# Neuroqueer Literacies in a Physics Context

## A Discussion on Changing the Physics Classroom Using a Neuroqueer Literacy Framework

Authors: Daniel P. Oleynik[1], Kiley Fridley[2], Liam G. McDermott[3]
1: University of Central Florida, 2: Ankeny Community School District, 3: University of Connecticut – Avery Point

**Introduction**

Neurodiversity, the paradigm shift away from a pathologization of cognitive difference and towards a celebration of cognitive diversity[1], in an educational context has important implications on how educators structure their classes and on how they develop inclusive and responsive pedagogies, which we discuss in this article. Being neurodivergent means having non-normative ways of thinking, sensing, and behaving[2]. Because of this, neurodivergent people learn and perform their understanding of physics differently than neurotypical people[3,4]. Kleekamp[5] and Smilges[6] call to our attention that when educators assess student competence, educators do so in a normative way. That is to say, educators privilege specific (non-disabled) ways of demonstrating understanding while pushing students who problem solve differently to the margins. By using a neuroqueer literacy framework[5,6] from Kleekamp and Smilges and applying ideas on neurodivergent physics learning and performance from McDermott[3], we provide recommendations for neuro-inclusive physics pedagogy.

Just as evaluating reading literacy constitutes the performance of reading – making meaning about a given text, physics literacy constitutes the performance of physics – making meaning about a given physical system. However, educators do not always assess physics literacy in ways which support all ways of learning and doing physics[7,8]. We bring in a neuroqueer literacy framework to inform our recommendations for neuro-inclusive pedagogy because, as we will discuss in the next section, neuroqueering involves questioning, subverting, and dismantling normative ways of learning and doing physics – something from which we believe that neuro-inclusive pedagogy benefits.

## Neuroqueer Literacy(ies)

Below, we provide definitions of some terms and definitions which may be new to readers:

1. Neurodivergence/Neurodivergent[2]
    a. An identity defined by non-normative ways of thinking, sensing, and behaving.
2. Queer[9,10]
    a. A spectrum of sexual orientation and gender identity outside of the normative ideas of the gender binary.
    b. To question, subvert, and dismantle normative and oppressive structures.
3. Neuroqueer[11,12]
    a. The intersection of queer and neurodivergent identity.
    b. To question, subvert, and dismantle structures to create a space which celebrates and encourages non-normative ways of thinking, sensing, and behaving.

Neuroqueer literacies is a framework which encourages educators to subvert, question, and subsequently dismantle how we define "literacy." Though Kleekamp[5] and Smilges[6], in their respective papers, focus on reading-specific literacy, we believe that the overarching concepts of neuroqueer literacies can *and should* be applied to physics. Neuroqueerness involves de-centering what we perceive as "the norm" – saying, *society expects me to be x, y, and z, but I do a, b, and c, and that's ok to be.* Thus, a goal of neuroqueering the classroom is to imagine ways in which doing a, b, and c are celebrated and encouraged, rather than forced to become x, y, and z. (For more reading on this idea, see: Remi Yergeau's "Authoring Autism"[12])

The "neuroqueer" in neuroqueer literacies connects aspects of queer theory and the experiences of queer students with neurodiversity and the experiences of neurodivergent students. Queerphobia and abelism are highly related, operating in similar ways, drawing on similar ideologies, and often co-occuring in practice. Thus, it is natural to connect them in a theoretical lens[13]. For example, the belief of queerness needing to be cured or fixed is almost identical to the ableist belief that disabilities and impairments should be cured within society[14]. As another example, pressure to stay within "the closet" around instructors exists for both neurodivergent and queer students[15,16]. Both queer and neurodivergent/disabled people are regularly expected to "pass" as straight, cisgender, neurotypical, or abled; and the default assumption is often that a person is both straight/cisgender and neurotypical/abled, unless they say otherwise (coming out of the closet) or assumptions based on appearance or behavior indicate otherwise (stereotyping).

Kleekamp's analysis of classroom observations, teacher reflections, and *in-situ* interviews elucidates three key themes of doing Neuroqueer Literacies: 1) *Presuming competence in neuroqueer literacy practices*, 2) *Asociality as a mode to produce countersocialities in neuroqueer literacy practices*, and 3) *Neuroqueer embodied invention in literacy practices*. These themes, which will be discussed and defined in the next section, are useful ways to subvert both the role that educators play in the classroom, and how educators understand what it means to generate knowledge in the classroom.

In addition, J. Logan Smilges[6] argues for a definition of literacy which takes into account the many varied ways which people engage in meaning-making. For instance, a neurodivergent person may take a calculated, iterative approach to applying different methods of solving a problem, which, to an outside viewer, could look like random application of formulae and equations[3]. Combining Smilges's ideas into Kleekamp's, our finalized themes and definitions of the neuroqueer literacies framework are:

1. Presuming competence in neuroqueer literacy practices,
    a. Students perform their competence in many varied ways; thus, educators should presume that students are competent, first and foremost
2. Asociality as a mode to produce countersocialities in neuroqueer literacy practices,
    a. Students learn and create knowledge in ways which go outside of expected social norms.
3. Neuroqueer embodied invention in literacy practices,
    a. Students solve problems through embodied movements.
4. Focus on sense-making above performance.
    a. Literacy should be defined as a student's ability to make meaning about a subject, rather than rote performance of tasks.

**Positioning within neuroqueer rhetoric**

We, as neurodivergent researchers and educators, view, construct, and share knowledge in ways that are counter to traditional (normative) "best"-practices[3,17]. Author Oleynik is a white, queer and neurodivergent physics education researcher who identifies with multiple impairments, including neurodivergence, chronic back pain, and being hard-of-hearing. Author Fridley is a white and neurodivergent high school physics educator who identifies with dyslexia, dyscalculia, and other neurodivergence. Author McDermott is a white, queer and neurodivergent physics education researcher, who identifies as multiply disabled, being both neurodivergent and hard-of-hearing.

We engage in neuroqueer physics literacies in our teaching and research, and our experience as neurodivergent students, educators, and researchers informs the recommendations in this article. As an example, being dyslexic greatly impacted Fridley's ability to consume written materials and her dyscalculia made it incredibly difficult to copy solutions to physics problems in a way that made sense. As a student who struggled with learning physics "the right way" and had to invent ways to make physics approachable for herself, she got into teaching with the goal of helping similar students. Fridley's goal as an educator is to inform and be informed about the best ways to provide physics education to all students regardless of ability and background.

The physics that Fridley teaches is a semester-long high school course designed to be the physics portion of a course for juniors and seniors. This course is a graduation requirement, and the class is designed for students who fall into one of the following categories: students who are not attending post-secondary education/not entering a STEM field, students who are not "science/math minded," or students who have difficulties in traditional classrooms. She sees many neurodivergent students, and about one-third of her students have an IEP or 504 plan[18] for various levels of ability. Some common things that she sees in many of her students are difficulty connecting ideas about physical phenomena (i.e. tying Newton's Second Law of Motion with existing knowledge about the way that objects move or interact), and limited success with traditional modes of assessment.

**Translating Neuroqueer Literacy to Scientific Literacy**

Historically[19], neurodivergent people have been assigned incapable of performing well in the classroom and thus assumed incompetent. This is because a neurotypical-normative education assumes that if a student is competent, then they will perform well in a task, therefore if a student performs well, they must be competent. Competence and performance are not necessarily bi-directional though. As an example, think about how many times you (neurodivergent or neurotypical) have made a brief glance at a sign, a billboard, or some other text, and initially assumed it said something wildly different than what it says in reality. Does mis-reading a sign mean you are not literate? Of course not. Having students understanding the physical processes which govern the universe and have the tools in place to "think like a physicist"[20] is critical to creating a more technologically and scientifically capable society. In physics, a student leaving an answer incomplete after applying multiple methods of problem solving (ie. An energy approach, a kinematics approach, a Newtonian approach) because they did not use the correct "trick" at the beginning of a problem does not necessarily make them physics illiterate.

In the table below, we break down the neuroqueer literacies framework into its constituent parts and provide recommendations for incorporating it into physics pedagogy. These recommendations are non-exhaustive and certainly warrant further study. Furthermore, these recommendations contain a decent amount of overlap with pedagogical frameworks such as Universal Design for Learning[21] and Culturally Relevant Pedagogy[22,23]; this list of recommendations is not meant to usurp these frameworks, but to add and to provide additional context. While using these recommendations as a guide, educators should maintain an open and honest dialogue with students to best support individual learning.

| Theme | Definition | Recommendations |
|---|---|---|
| **Presuming competence in neuroqueer literacy practices** | Students perform their competence in many varied ways; thus, educators should presume that students are competent, first and foremost. | Give feedback in a way which points students to the logical basis of their alternative conceptions. |
| | | Experiment with alternative grading such as ungrading or mastery-based grading.<br><br>Use a positive scoring rubric that does not take off points for extra work. |
| **Asociality as a mode to produce countersocialities in neuroqueer literacy practices** | Students learn and create knowledge in ways which go outside of expected social norms. | Remove strict attendance requirements.<br><br>Support learner autonomy in research/lab experiences.<br><br>Allow students to work alone/choose partners when possible<br><br>Have flexible seating arrangements.<br><br>Do not ask students to justify why they think differently, including forcing them to disclose or educate you about a diagnosis<br><br>Include breaks within long lectures (once every hour).<br><br>Do not pressure students to "come out of the closet" or "unmask," even with the best intentions. |
| **Neuroqueer embodied invention in literacy practices** | Students solve problems through embodied movements. | Allow/encourage students use their bodies when demonstrating new ideas.<br><br>Engage students with multisensory representations of phenomena.<br><br>Engage students with multisensory and hands-on learning opportunities and demonstrations that |

| | | encourage student to think in words and in images. |
|---|---|---|
| **Focus on sense-making above performance.** | Literacy should be defined as a student's ability to make meaning about a subject, rather than rote performance of tasks. | Provide multiple modes of assessment.<br><br>Prompt metacognitive self-assessment for students to reflect on levels of learning.<br><br>Provide materials (ie., slides, notes) in advance.<br><br>Explicitly describe expectations for coursework in rubrics.<br><br>Focus on processes when grading rather than the correct numerical answer.<br><br>Design flexible rubrics that allow for full points via multiple solution paths. |

Table 1. Definitions of the four neuroqueer literacies themes and recommendations for physics pedagogy which stem from the themes.

From these four themes, we can build classroom practices which support all students regardless of ability or other marginalized status. It is important that instructors, when constructing pedagogies for neuroqueer literacies, that structures are constructed proactively, and not reactively. Furthermore, we call on educators when applying our recommendations to interrogate the purpose of their class and interpret our recommendations accordingly. These recommendations are flexible. For instance, it may be a crucial point of a laboratory class that a student has perfect attendance. Therefore, our recommendation for attendance policies may be better interpreted as allowing make-up labs throughout the week. It should also be noted that each recommendation is not specific to neuroqueer/neurodivergent students, nor is it in any way the end-all-be-all of recommendations for neuroqueer literacy pedagogy. Instructors may not have the answers to every student's situation and that is ok. What is important is that instructors engage students with compassion. There is still much left to learn about the ways neurodivergent learners learn and perform physics. However, with these recommendations as a jumping-off point, we believe that educators can, at the very least, begin the process of dismantling ableist practices and structures in the classroom.

**Acknowledgements**

This work was funded by the National Science Foundation STEM Education Individual Postdoctoral Research Fellowship (STEMEdIPRF) (#2411711).

**Author Bios:**

Daniel Oleynik: Daniel Oleynik is a current graduate research assistant in their last year at the University of Central Florida His work focuses on improving the access, accommodations and environment for students with disabilities. He does this by analyzing the attitudes and beliefs of practicing physics to create suggestions for instructors and mentors to apply within their classrooms.

Kiley Fridley: Kiley Fridley is a high school physics teacher for Ankeny Community School District. She has a bachelor's in physics and a master's in teaching from Iowa State University. Her goal is to make physics accessible to students of all backgrounds and abilities.

Liam McDermott, Ph.D (vtb24001@uconn.edu): Dr. McDermott is a postdoctoral fellow at the University of Connecticut - Avery Point. His work focuses on investigating how neurodivergent physics students learn and perform physics. He intends to use the results of his research to develop neuro-inclusive pedagogy to support all students.

---

[1] N. Walker (2016). *Autism & the pathology paradigm*. Neuroqueer - The Writings of Dr. Nick Walker. https://neuroqueer.com/autism-and-the-pathology-paradigm/

[2] N. Walker (2014). *Neurodiversity: Some basic terms & definitions*. Neuroqueer - The Writings of Dr. Nick Walker. https://neuroqueer.com/neurodiversity-terms-and-definitions/

[3] L. McDermott, (2025). Neuroqueering Physics Literacy. Journal for Theoretical & Marginal Mathematics Education, 3(2), Article 0304. https://doi.org/10.5281/zenodo.14888967

[4] S. Salvatore, C. White, & S. Podowitz-Thomas, “Not a cookie cutter situation”: how neurodivergent students experience group work in their STEM courses. *IJ STEM Ed* **11**, 47 (2024). https://doi.org/10.1186/s40594-024-00508-0

[5] M. C. Kleekamp, (2020). “No! Turn the pages!” Repositioning neuroqueer literacies. *Journal of Literacy Research*, *52*(2), 113-135. https://doi.org/10.1177/1086296X20915531

[6] J. L. Smilges, (2021). Neuroqueer literacies; or, against able-reading. *College Composition & Communication*, *73*(1), 103-125. https://doi.org/10.58680/ccc202131589

[7] J. Schmidt, (1990). Physics literacy. *Physics Today*, *43*(11), 60-67. https://doi.org/10.1063/1.881217

[8] A. Hobson, (2003). Physics literacy, energy and the environment. *Physics Education*, *38*(2), 109. DOI: 10.1088/0031-9120/38/2/301

[9] N. Sullivan, (2003). *A critical introduction to queer theory*. NYU Press.

[10] J. Butler, (2020). Critically queer. In *Playing with fire* (pp. 11-29). Routledge.

[11] N. Walker, (2015). *Neuroqueer: An Introduction*. Neuroqueer - The Writings of Dr. Nick Walker. https://neuroqueer.com/neuroqueer-an-introduction/

[12] M. R. Yergeau, (2018). *Authoring autism: On rhetoric and neurological queerness*. Duke university press.

[13] McRuer, R. (2008). Crip theory. Cultural signs of queerness and disability.

[14] Bailey, C. W. (2019). On the impossible: Disability studies, queer theory, and the surviving crip. Disability Studies Quarterly, 39(4). https://doi.org/10.18061/dsq.v39i4.6580

[15] Allen, L., Cowie, L., & Fenaughty, J. (2020). Safe but not safe: LGBTTIQA+ students’ experiences of a university campus. *Higher Education Research & Development*, *39*(6), 1075-1090. https://doi.org/10.1080/07294360.2019.1706453

[16] Lierman, A. (2024). *The Struggle You Can’t See: Experiences of Neurodivergent and Invisibly Disabled Students in Higher Education* (p. 284). Open Book Publishers.

[17] J. P. Barnett, (2024). Neuroqueer frontiers: Neurodiversity, gender, and the (a) social self. *Sociology Compass*, *18*(6), e13234. https://doi.org/10.1111/soc4.13234

[18] J. Rawe, (2024). The difference between IEPs and 504 plans. Understood. https://www.understood.org/en/articles/the-difference-between-ieps-and-504-plans

[19] S. M. Acevedo, & E. A. Nusbaum, (2020). “Autism, neurodiversity, and inclusive education.” Oxford Research Encyclopedia of Education.

[20] A. Van Heuvelen, (1991). Learning to think like a physicist: A review of research-based instructional strategies. *American Journal of physics*, *59*(10), 891-897. https://doi.org/10.1119/1.16667

[21] CAST (2024). Universal Design for Learning Guidelines version 2.0. Wakefield, MA: Author.

[22] Brown-Jeffy, S., & Cooper, J. E. (2011). Toward a conceptual framework of culturally relevant pedagogy: An overview of the conceptual and theoretical literature. *Teacher education quarterly*, *38*(1), 65-84.

[23] Ladson-Billings, G. (1995). Toward a theory of culturally relevant pedagogy. *American educational research journal*, *32*(3), 465-491. https://doi.org/10.3102/00028312032003465